\documentclass[twocolumn,english,superscriptaddress,longbibliography]{revtex4-1}
\usepackage[T1]{fontenc}
\usepackage[utf8]{inputenc}
\usepackage{verbatim}
\usepackage{bm}
\usepackage{amsmath}
\usepackage{amssymb}
\usepackage{graphicx}

\makeatletter
\usepackage{braket}
\usepackage{CJK}
\usepackage{bm}
\usepackage{tikz}
\usepackage{pgfplots}
\usepackage{pgf}
\usepackage{subfigure}

\makeatother

\usepackage{babel}
\begin{document}
\title{Asymmetry of the Geometrical Resonances of Composite Fermions}
\author{Guangyue Ji}
\affiliation{International Center for Quantum Materials, Peking University, Beijing
100871, China}
\author{Junren Shi}
\email{junrenshi@pku.edu.cn}

\affiliation{International Center for Quantum Materials, Peking University, Beijing
100871, China}
\affiliation{Collaborative Innovation Center of Quantum Matter, Beijing 100871,
China}
\begin{abstract}
We propose an experiment to test the uniform-Berry-curvature picture
of composite fermions. We show that the asymmetry of geometrical resonances
observed in a periodically modulated composite fermion system can
be explained with the uniform-Berry-curvature picture. Moreover, we
show that an alternative way of modulating the system, i.e., modulating
the external magnetic field, will induce an asymmetry opposite to
that of the usual periodic grating modulation which effectively modulates
the Chern-Simons field. The experiment can serve as a critical test
of the uniform-Berry-curvature picture, and probe the dipole structure
of composite fermions proposed by Read.
\end{abstract}
\maketitle

\section{Introduction}

A two-dimensional electron system (2DES) subjected to a strong perpendicular
magnetic field exhibits exotic many-body states, in particular, the
fractional quantum Hall states at odd-denominator filling factors$\,$\citep{tsui_two-dimensional_1982,jain_composite-fermion_1989}
and the Fermi-liquid-like states at even-denominator fillings$\,$\citep{kalmeyer_metallic_1992,halperin_theory_1993}.
Jain's composite fermion (CF) theory provides a unified understanding
to these states$\,$\citep{jain2007composite}. A CF can be regarded
as an electron attached with $2p$ quantum vortices and feels an effective
magnetic field $B^{*}=B-b_{2p}^{\textrm{CS}}$ with $B$ being the
external magnetic field and $b_{2p}^{\textrm{CS}}=2pn_{\textrm{e}}\phi_{0}$
the emergent Chern-Simons (CS) field, where $n_{\textrm{e}}$ is the
density of electrons and $\phi_{0}=h/e$ is the quanta of magnetic
flux$\,$\citep{jain_composite-fermion_1989,halperin_theory_1993}.
The Halperin-Lee-Read (HLR) theory treats the CF as an electron-like
particle and predicts that CFs form a Fermi liquid at even-dominator
filling $\nu=1/2p$ for which the effective magnetic field $B^{*}=0$$\,$\citep{halperin_theory_1993}.
The Fermi liquid state is confirmed by various experiments$\,$\citep{Heinonen}.
Though the HLR theory achieves great successes in explaining various
observed phenomena, it does not predict a correct CF Hall conductivity
$\sigma_{xy}^{\textrm{CF}}=-e^{2}/2h$ at half-filling as required
by the particle-hole symmetry$\,$\citep{kivelson_composite-fermion_1997}.
Motivated by the difficulty, Son proposes that the CF is a Dirac particle$\,$\citep{son_is_2015}.
In the Dirac theory, the CF is considered as a vortex dual of a Dirac
electron coupling to an emergent gauge field. However, its microscopic
basis is not yet clarified$\,$\citep{balram_nature_2016}. On the
other hand, Shi et al. derive the dynamics of the CF Wigner crystal
from the microscopic Rezayi-Read wave function and find that CFs are
subjected to a Berry curvature uniformly distributed in momentum space$\,$\citep{shi_dynamics_2018}.
Based on that, they propose the uniform-Berry-curvature picture of
CFs$\,$\citep{shi2017chernsimons}. A calculation of the Berry phase
of CFs from a microscopic wave function by Geraedts et al. seems to
lend a support to the Dirac picture$\,$\citep{geraedts_berry_2018}.
However, a refined calculation suggests otherwise$\,$\citep{ji2019berry}.
Actually, the Berry curvature is analytically shown to be uniform
for the Rezayi-Read wave function$\,$\citep{rezayi_fermi-liquid-like_1994}.
Although the two pictures look quite different, both predict that
a CF accumulates a $\pi$ Berry phase when it moves around the Fermi
circle.

The manifestations of the $\pi$ Berry phase have been observed in
a number of experiments and numerical calculations. In the numerical
simulations of the infinite-cylinder density matrix renormalization
group, the suppression of $2k_{\textrm{F}}$ backscattering off particle-hole
symmetric impurities is interpreted as a result of the $\pi$ Berry
phase$\,$\citep{geraedts_half-filled_2016}. In the Shubnikov--de
Haas oscillation experiments of CFs at a fixed magnetic field, the
$\pi$ Berry phase is shown to appear in the magnetoresistivity formula$\,$\citep{pan_berry_2017}.
In the geometrical resonance experiments of CFs with periodic grating
modulations, the asymmetry of the commensurability condition on the
two sides about half filling observed in Ref.$\,$\citep{kamburov_what_2014}
can also be explained as a result of the $\pi$ Berry phase (see below).
Though these studies convincingly show the presence of the $\pi$
Berry phase, they can not differentiate the Dirac picture and the
uniform-Berry-curvature picture.

In this paper, we propose an experiment to test the uniform-Berry-curvature
picture. First, we show that the uniform-Berry-curvature picture predicts
a Fermi wave vector different from the HLR theory but same as the
Dirac theory$\,$\citep{son_is_2015}. The asymmetry of the commensurability
conditions in Ref.$\,$\citep{kamburov_what_2014} can be explained
with the modified Fermi wave vector. Next, we show that the uniform-Berry-curvature
picture is equivalent to the dipole picture initially proposed by
Read$\,$\citep{read_theory_1994,read_recent_1996}. In the dipole
picture, it becomes obvious that the external magnetic field $\bm{B}$
and the CS field $\bm{b}^{\textrm{CS}}$ are coupling to different
internal degrees of freedom in a CF, i.e., the electron and the quantum
vortices, respectively (see Figure.$\,$\ref{figure1_1}). We show
that a geometrical resonance experiment with a periodically modulated
external magnetic field will yield an asymmetry opposite to that of
the usual periodic grating modulation. This experiment can serve as
a critical test to the uniform-Berry-curvature picture, and at the
same time, probe the ``subatomic'' dipole structure of CFs.

The remainder of the paper is organized as follows. In Sec.$\,$\ref{sec:Uniform-Berry-curvature-picture-},
we derive the Fermi wave vector based on the uniform-Berry-curvature
picture, and show that the uniform-Berry-curvature picture is equivalent
to the dipole picture. In Sec.$\,$\ref{sec:Periodic-scalar-potential},
we study the periodic scalar potential modulation of CFs. In Sec.$\,$\ref{sec:Periodic-external-magnetic},
we study the periodic external magnetic field modulation of CFs. In
Sec.$\,$\ref{sec:Discussions-and-summary}, we discuss and summarize
our results.

\section{Uniform-Berry-curvature picture and dipole picture\label{sec:Uniform-Berry-curvature-picture-}}

In the uniform-Berry-curvature picture, the equations of motion (EOMs)
of CFs read
\begin{align}
\dot{\bm{x}} & =\frac{\bm{p}}{m_{\textrm{CF}}^{*}}+\frac{1}{eB}\hat{z}\times\dot{\bm{p}},\label{eq:EOM1_1}\\
\dot{\bm{p}} & =-eB^{*}\hat{z}\times\dot{\bm{x}},\label{eq:EOM1_2}
\end{align}
where $\bm{x}$, $\bm{p}$, and $m_{\textrm{CF}}^{*}$ are the position,
momentum, and effective mass of a CF, respectively$\,$\citep{shi_dynamics_2018}.
A distinctive feature of the uniform-Berry-curvature picture is the
presence of a uniform Berry curvature in the momentum space, which
is not presented in the conventional HLR theory$\,$\citep{shi_dynamics_2018}.
As a result, it predicts a Fermi wave vector different from the HLR
theory. In the HLR theory, the CF is treated as an electron-like particle.
It predicts a Fermi wave vector $k_{\textrm{F}}=\sqrt{4\pi n_{\textrm{e}}}$.
On the other hand, in the uniform-Berry-curvature picture, due to
the presence of the Berry curvature $\Omega_{z}=1/eB$ in Eq.$\,$(\ref{eq:EOM1_1}),
the phase-space density of states is modified by a factor $D=1-B^{*}/B$$\,$\citep{xiao_berry_2005}.
The Fermi wave vector $k_{\textrm{F}}$ of CF can be determined through
the condition $\pi k_{\textrm{F}}^{2}D/\left(2\pi\right)^{2}=n_{\textrm{e}}$,
and is
\begin{align}
k_{\textrm{F}} & =\sqrt{\frac{eB}{\hbar}},\label{eq:wavevector}
\end{align}
which is different from the prediction of the HLR theory and independent
of $n_{\textrm{e}}$. This result is the same as the Dirac theory.
The coincidence is not surprising because both the pictures have a
$\pi$-Berry phase along the Fermi circle. To differentiate the two
pictures, one has to probe deeper.

\begin{figure}[t]
\centering\subfigure[]{\label{figure1_1}\includegraphics[width=6cm]{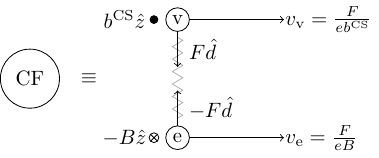}}\hfill{}\subfigure[$B>b^{\text{CS}}$]{\label{figure1_2}\includegraphics[width=3.5cm]{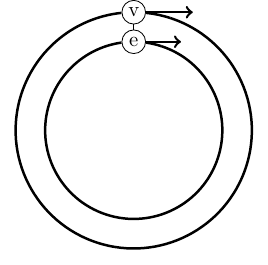}}\hfill{}\hfill{}\subfigure[$B<b^{\text{CS}}$]{\label{figure1_3}\includegraphics[width=3.5cm]{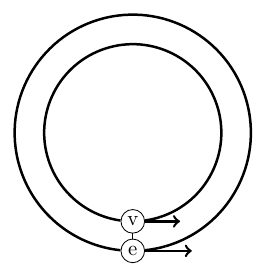}}

\caption{\label{figure1} Cyclotron orbits of a CF under various conditions:
(a) The dipole structure of a CF: a CF consists of an electron (e)
and two quantum vortices (v). They are bounded together by a mutual
central force $F\propto\left|\bm{d}\right|$. e is coupled to the
external magnetic field $-B\hat{z}$, and v is coupled to the Chern-Simons
field $b^{\textrm{CS}}\hat{z}$. When $B=b^{\text{CS}}$, e and v
have the same velocity $v=(-F)/(-eB)=F/eb^{\textrm{CS}}$ and move
linearly; (b) When $B>b^{\text{CS}}$ , e and v have different velocities,
resulting in a cyclotron motion. Because v is faster than e, the cyclotron
radius of v is larger than that of e, i.e., $R_{\mathrm{c}}^{(\mathrm{v})}\equiv R_{\textrm{c}}^{*}>R_{\textrm{c}}^{\left(\textrm{e}\right)}$;
(c) When $B<b^{\text{CS}}$, the opposite is true, i.e., $R_{\textrm{c}}^{\left(\textrm{e}\right)}>R_{\textrm{c}}^{(\mathrm{v})}$.
The asymmetry between (b) and (c) is responsible for the asymmetry
observed in geometrical resonance experiments. By using the usual
grating modulation, one measures $R_{\mathrm{c}}^{(\mathrm{v})}$.
By using the magnetic field modulation, on the other hand, one measures
$R_{\textrm{c}}^{\left(\textrm{e}\right)}$. It is obvious that the
two different approaches will yield opposite asymmetries.}
\end{figure}

The uniform-Berry-curvature picture is actually equivalent to the
dipole picture initially proposed by Read$\,$\citep{read_theory_1994,read_recent_1996}.
To see that, we can rewrite the EOMs with the new variables $\bm{x}^{\textrm{v}}\equiv\bm{x}$
and $\bm{x}^{\textrm{e}}=\bm{x}^{\textrm{v}}-\hat{z}\times\bm{p}/eB$:
\begin{align}
-eB\hat{z}\times\dot{\bm{x}}^{\textrm{e}} & =\frac{\partial\varepsilon}{\partial\bm{x}^{\textrm{e}}},\\
eb^{\textrm{CS}}\hat{z}\times\dot{\bm{x}}^{\textrm{v}} & =\frac{\partial\varepsilon}{\partial\bm{x}^{\textrm{v}}},
\end{align}
where $\bm{x}^{\textrm{e}}$ and $\bm{x}^{\textrm{v}}$ are interpreted
as the position of the electron and quantum vortices in a CF, respectively,
and $\varepsilon\propto\left|\bm{x}^{\textrm{e}}-\bm{x}^{\textrm{v}}\right|^{2}$
is the binding energy between the electron and the quantum vortices$\,$\citep{shi_dynamics_2018}.
The momentum $\bm{p}$ of a CF is interpreted as $\bm{p}=eB\hat{z}\times\bm{d}$
with the displacement $\bm{d}\equiv\bm{x}^{\textrm{e}}-\bm{x}^{\textrm{v}}$
(see Figure.$\,$\ref{figure1_1}). From the new form of the EOMs,
it is clear that the electron is only coupled to the external electromagnetic
field $\bm{B}$ while the quantum vortices are only coupled to the
emergent CS field $\bm{b}^{\textrm{CS}}$. Moving a CF in the momentum
space is equivalent to fixing the quantum vortices and moving the
electron in the real space. The Aharonov-Bohm phase accumulated by
the electron is nothing but the Berry phase expected from the uniform
Berry curvature in Eq.$\,$(\ref{eq:EOM1_1})$\,$\citep{shi2017chernsimons}.
It also becomes obvious that the external magnetic field and the CS
field are not equivalent microscopically since they are coupling to
different internal degrees of freedom. Therefore, we anticipate that
modulating the external magnetic field $\bm{B}$ and the CS field
$\bm{b}^{\textrm{CS}}$ have different effects on CFs.

\section{Scalar potential modulation\label{sec:Periodic-scalar-potential}}

Weiss et al. show that when a 2DES is weakly modulated by a one dimensional
periodic scalar potential, its magnetoresistance shows an oscillation
with respect to $2R_{\textrm{c}}/a$, where $R_{\textrm{c}}$ is the
cyclotron radius and $a$ is the period of the modulation$\,$\citep{weiss_magnetoresistance_1989}.
When a 2DES is at an even-dominator filling factor $\nu=1/2p$, CFs
feel a zero effective magnetic field $B^{*}=0$ . It is nature to
expect that the Weiss oscillation can also be observed in CF systems
when the effective magnetic field deviates from zero. This has been
confirmed by a number of geometrical resonance experiments for CFs$\,$\citep{willett_experimental_1993,kang_how_1993,goldman_detection_1994,smet_magnetic_1996}.
In experiments, the scalar potential modulation is achieved by imposing
a grating pattern$\,$\citep{skuras_anisotropic_1997}.

For CF systems, a periodic scalar potential modulation is equivalent
to a modulation of the CS field for CFs. In such a modulation, CFs
are subjected to a weak electrostatic potential modulation $\delta V^{\textrm{ext}}\left(x\right)=V^{\textrm{ext}}\cos(2\pi x/a)$.
The electrostatic potential will induce a modulation of the electron
density $\delta n_{\textrm{e}}$, which in turn induces a modulation
of the CS field $\delta b^{\textrm{CS}}=2\phi_{0}\delta n_{\textrm{e}}$.
The energy corrections associated with the electrostatic potential
and the CS field are $-e\delta V^{\textrm{ext}}$ and $-e\dot{\bm{x}}\cdot\delta\bm{a}^{\textrm{CS}}$,
respectively. By assuming a non-interacting CF model, the ratio of
these two contributions is $\pi/ak_{\textrm{F}}\ll1$ (e.g., for $B=14\,\textrm{T}$
and $a=200\,\textrm{nm}$, $\pi/ak_{\textrm{F}}\approx0.1$)$\,$\citep{peeters_quantum_1993}.
As a result, the effect of the CS field modulation dominates in this
case.

The commensurability condition can be derived semi-classically as
shown in Refs.$\,$\citep{beenakker_guiding-center-drift_1989,peeters_quantum_1993},
in which the modulation is treated as a perturbation. In the absence
of the modulation, for a CF on the Fermi circle, the solutions of
Eqs.$\,$(\ref{eq:EOM1_1}, \ref{eq:EOM1_2}) are
\begin{align}
\bm{x}\left(t\right) & =\bm{x}_{0}+R_{\textrm{c}}^{*}\left[-\cos\left(\omega_{\textrm{c}}^{*}t+\varphi\right),\sin\left(\omega_{\textrm{c}}^{*}t+\varphi\right)\right],\label{eq:x}\\
\bm{p}\left(t\right) & =\hbar k_{\textrm{F}}\left[\sin\left(\omega_{\textrm{c}}^{*}t+\varphi\right),\cos\left(\omega_{\textrm{c}}^{*}t+\varphi\right)\right],\label{eq:p}
\end{align}
where $\bm{x}_{0}$, $R_{\textrm{c}}^{*}=\hbar k_{\textrm{F}}/eB^{*}$
and $\omega_{\textrm{c}}^{*}=eB^{*}/Dm_{\textrm{CF}}^{*}$ are the
center coordinate, radius and frequency of the cyclotron orbit, respectively,
and $\varphi$ is a phase factor. Without the periodic modulation,
all orbits have a degenerate energy. In the presence of the weak periodic
modulation, the degeneracy is split. The correction to the energy,
to the first order, is the average energy change due to the CS field
modulation during a period of the cyclotron motion $T=2\pi/\omega_{\textrm{c}}^{*}$:
\begin{align}
\delta U & \approx\frac{1}{T}\int_{0}^{T}dt(-e\dot{\bm{x}}\cdot\delta\bm{a}^{\textrm{CS}})\nonumber \\
 & =(2ek_{\textrm{F}}V^{\textrm{ext}}/q)J_{1}(qR_{\textrm{c}}^{*})\cos qx_{0}
\end{align}
 where $q=2\pi/a$ and $J_{1}\left(x\right)$ is the first Bessel
function$\,$\citep{peeters_quantum_1993}. In the weak effective
magnetic field limit $qR_{\textrm{c}}^{*}\gg1$, $\delta U\approx-\sqrt{2/\pi qR_{\textrm{c}}^{*}}(2ek_{\textrm{F}}V^{\textrm{ext}}/q)\cos qx_{0}\cos(qR_{\textrm{c}}^{*}+\pi/4)$.
The energy correction depends on the center position $x_{0}$, resulting
in the broadening of the Landau level. The broadening caused by the
modulation is proportional to $\cos\left(qR_{\textrm{c}}^{*}+\pi/4\right)$,
which vanishes when the commensurability condition $2R_{\textrm{c}}^{*}/a=i+\gamma$
with $\gamma=1/4$ is fulfilled. One may assume that the conductivity
along the direction transverse to the modulation is proportional to
the broadening$\,$\citep{peeters_electrical_1992,peeters_quantum_1993}.
As a result, the commensurability condition is manifested in experiments
as a series of the minimum of the longitudinal magnetoresistance.
For the Fermi wave vector shown in Eq.$\,$(\ref{eq:wavevector}),
the commensurability condition can be written as:
\begin{align}
\frac{B_{0}}{\left|B_{i}^{*}\right|} & \approx\frac{a}{2}\sqrt{\frac{eB_{0}}{\hbar}}\left(i+\gamma\right)+\left\{ \begin{array}{cc}
-\frac{1}{2} & B^{*}>0\\
\frac{1}{2} & B^{*}<0
\end{array}\right.\label{eq:condition1}
\end{align}
for $\left|B_{i}^{*}\right|\ll B_{0}$, where $B_{0}\equiv2n_{\textrm{e}}\phi_{0}$
is the magnetic field at the half-filling, and $B_{i}^{*}$ is the
effective magnetic field of the $i$-th magnetoresistance minima.
We see that the commensurability condition shows an asymmetry between
the particle $\left(B^{*}>0\right)$ and hole $\left(B^{*}<0\right)$.

\begin{figure}[t]
\includegraphics[scale=0.99]{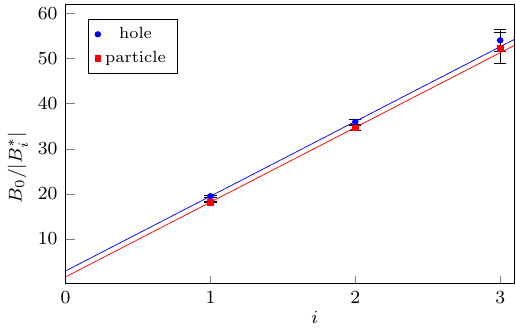}\caption{\label{figure2} $B_{0}/\left|B_{i}^{*}\right|$ versus $i$. The
circles (squares) are data for hole (particle) and the error bar is
also shown. The slope of both lines is $16.61\pm0.11$. The difference
of the intercepts of two lines is $-1.33\pm0.39$. The fitting value
of phase factor is $\gamma=0.13\pm0.01$. Points are data adapted
from Ref.$\,$\citep{kamburov_what_2014} with $B_{0}=14.383\,\textrm{T}$.}
\end{figure}

The asymmetry had actually been observed in experiments. We adapt
and fit the experimental results of Ref.$\,$\citep{kamburov_what_2014},
and show them in Figure.$\,$\ref{figure2}. One can see that for
all index $i$'s, the value of $B_{0}/\left|B_{i}^{*}\right|$ with
$B_{i}^{*}<0$ (hole) sits above that with $B_{i}^{*}>0$ (particle).
The vertical shift of the two lines is $\Delta\left(B_{0}/\left|B_{i}^{*}\right|\right)=-1.33\pm0.39$,
close to the prediction $\Delta\left(B_{0}/\left|B_{i}^{*}\right|\right)=-1$.

\section{External magnetic field modulation\label{sec:Periodic-external-magnetic}}

In this section, we show that a weak periodic modulation of the external
magnetic field will induce an asymmetry opposite to that of the periodic
scalar potential modulation. First, we derive the commensurability
condition with respect to the external magnetic field modulation.
Then, we consider the effect of induced density modulation and determine
when the unwanted density modulation effect can be suppressed.

\subsection{Direct modulation effect}

When a weak periodic modulation of the external magnetic field $\delta\bm{B}\left(\bm{x}\right)=\delta B\cos\bm{q}\cdot\bm{x}\hat{z}$
is applied to a 2DES, it couples to the electron in the CF and the
energy correction is $\delta U=-e\dot{\bm{x}}^{\textrm{e}}\cdot\delta\bm{A}\left(\bm{x}^{\textrm{e}}\right)$
with $\delta\bm{A}\left(\bm{x}\right)$ being the vector potential
with respect to $\delta\bm{B}\left(\bm{x}\right)$. Note that $\delta U$
in the current case is related to the electron coordinate $\bm{x}^{\textrm{e}}$
instead of $\bm{x}$ as in the previous case. We can determine the
commensurability condition just as we do in the last section. In the
absence of the modulation, from Eqs.$\,$(\ref{eq:x}, \ref{eq:p}),
we determine 
\begin{align}
\bm{x}^{\textrm{e}}\left(t\right) & =\bm{x}\left(t\right)-\hat{z}\times\bm{p}\left(t\right)/eB\nonumber \\
 & =\bm{x}_{0}+R_{\textrm{c}}^{\left(\textrm{e}\right)}\left[-\cos\left(\omega_{\textrm{c}}^{*}t+\varphi\right),\sin\left(\omega_{\textrm{c}}^{*}t+\varphi\right)\right],
\end{align}
 with
\begin{equation}
R_{\textrm{c}}^{\left(\textrm{e}\right)}=DR_{\textrm{c}}^{*}.
\end{equation}
Therefore, the electron has a cyclotron radius different from that
of the quantum vortices. In this case, the average energy change of
a CF due to the external magnetic field modulation during a period
of the cyclotron motion is
\begin{align}
\delta U & \approx\frac{1}{T}\int_{0}^{T}dt(-e\dot{\bm{x}}^{\textrm{e}}\cdot\delta\bm{A}(\bm{x}^{\textrm{e}}))\nonumber \\
 & =(e\omega_{\textrm{c}}^{*}R_{\textrm{c}}^{(\textrm{e})}\delta B/q)J_{1}(qR_{\textrm{c}}^{\left(\textrm{e}\right)})\cos(qx_{0}).
\end{align}
 In the weak effective magnetic field limit $qR_{\textrm{c}}^{\left(\textrm{e}\right)}\gg1$,
$\delta U\approx-\sqrt{2/\pi qR_{\textrm{c}}^{\left(\textrm{e}\right)}}(e\omega_{\textrm{c}}^{*}R_{\textrm{c}}^{\left(\textrm{e}\right)}\delta B/q)\cos(qx_{0})\cos(qR_{\textrm{c}}^{\left(\textrm{e}\right)}+\pi/4)$.

As a result, the commensurability condition becomes $2R_{\textrm{c}}^{\left(\textrm{e}\right)}/a=i+\gamma$,
and can be written as:
\begin{align}
\frac{B_{0}}{\left|B_{i}^{*}\right|} & \approx\frac{a}{2}\sqrt{\frac{eB_{0}}{\hbar}}\left(i+\gamma\right)+\left\{ \begin{array}{cc}
\frac{1}{2} & B^{*}>0\\
-\frac{1}{2} & B^{*}<0
\end{array}\right..\label{eq:condition2}
\end{align}
for $\left|B_{i}^{*}\right|\ll B_{0}$. We see that the value of $B_{0}/\left|B_{i}^{*}\right|$
with $B_{i}^{*}>0$ (electron) now sits above that with $B_{i}^{*}<0$
(hole). The asymmetry is opposite to the asymmetry induced by the
CS field modulation.

Based on the result, we propose a new geometrical resonance experiment
with a modulating external magnetic field. The inverse of the asymmetry
would be the signature confirming the underlying ``subatomic'' structure
of the CF.

\subsection{Induced CS field modulation}

However, there is still a complexity for the proposed experiment.
This is because the energy of the lowest Landau level (LLL) is proportional
to $B$, and the modulation of $B$ will introduce a modulation of
the effective potential felt by CFs$\,$\citep{simon_composite_1996}.
While the direct effect of the effective potential is negligible,
the CS field induced by modulation may not be small. To estimate the
modulation amplitude of the induced CS field, we apply the density
functional approach$\,$\citep{ferconi_edge_1995,heinonen_ensemble_1995,zhao_density-functional_2017}.
By ignoring the effect of density gradient, the grand canonical energy
functional $\mathcal{E}$ of the system can be approximated as:
\begin{multline}
\mathcal{E}\left[n\right]=\int d\bm{r}\left[-\mu n\left(\bm{r}\right)+\left(\frac{\hbar e}{2m_{\textrm{b}}}+\frac{g\mu_{\textrm{B}}}{2}\right)B\left(\bm{r}\right)n\left(\bm{r}\right)\right.\\
\left.+v_{\textrm{xc}}\left[n\left(\bm{r}\right)\right]n\left(\bm{r}\right)+\frac{e^{2}}{8\pi\epsilon}\int d\bm{r}'\frac{\Delta n\left(\bm{r}\right)\Delta n\left(\bm{r}'\right)}{\left|\bm{r}-\bm{r}'\right|}\right]
\end{multline}
where $\mu$ is the chemical potential, the second term is the kinetic
and Zeeman energy of electrons in the LLL with $m_{\textrm{b}}$,
$g$ and $\mu_{\textrm{B}}$ being the band mass, the effective Land$\acute{\textrm{e}}$
factor and the Bohr magneton, respectively, $v_{\textrm{xc}}\left[n\left(\bm{r}\right)\right]$
is the exchange-correlation energy per particle, the last term is
the Coulomb energy due to the density modulation $\Delta n\left(\bm{r}\right)$.%
{} We adopt the interpolation formula of the exchange-correlation energy
presented in Ref.$\,$\citep{fano_interpolation_1988}, which is a
function of the filling factor $\nu=n/(eB/h)$. Under the local density
approximation, the exchange-correlation energy can be written as:
\begin{equation}
v_{\textrm{xc}}\left[n\left(\bm{r}\right)\right]=(e^{2}/4\pi\epsilon l_{B}(\bm{r}))u\left(\nu(\bm{r})\right),\label{eq:exchange-correlation energy}
\end{equation}
 with 
\begin{align}
u\left(\nu\right)= & -(\pi/8)^{1/2}\nu-0.782\nu^{1/2}(1-\nu)^{3/2}\nonumber \\
 & +0.683\nu(1-\nu)^{2}-0.806\nu^{3/2}(1-\nu)^{5/2},
\end{align}
where $l_{B}\equiv\sqrt{\hbar/eB(\bm{r})}$ is the magnetic length
and $\epsilon$ is the static permittivity$\,$\citep{fano_interpolation_1988}.

To determine the density modulation due to the modulation of the external
magnetic field, we minimize the energy functional with respect to
the density, and obtain
\begin{align}
\mu= & \left(\frac{\hbar e}{2m_{\textrm{b}}}+\frac{g\mu_{\textrm{B}}}{2}\right)B\left(\bm{r}\right)+\left.\frac{e^{2}}{4\pi\epsilon l_{B}(\bm{r})}\frac{\partial[u\left(\nu\right)\nu]}{\partial\nu}\right|_{\nu=\frac{n(\bm{r})}{eB(\bm{r})/h}}\nonumber \\
 & +\frac{e^{2}}{4\pi\epsilon}\int d\bm{r}'\frac{\Delta n\left(\bm{r}'\right)}{\left|\bm{r}-\bm{r}'\right|}.
\end{align}
For a weak periodic modulation of the magnetic field $B(\bm{r})=B+\delta B(\bm{r})$
and $\delta B(\bm{r})=\delta B\cos(2\pi x/a)$, we have $\Delta n(\bm{r})\approx\delta n\cos(2\pi x/a)$.
By assuming that both $\delta n$ and $\delta B$ are small quantities,
it is easy to obtain
\begin{align}
\frac{\delta n}{n} & \approx-\frac{a_{\textrm{c}}/l_{B}-\beta_{1}}{a/l_{B}-\beta_{2}}\frac{\delta B}{B}
\end{align}
where $a_{\textrm{c}}\equiv2\pi\left(1+gm_{\textrm{b}}/2m_{\textrm{e}}\right)a_{\textrm{B}}^{\ast}$
with $a_{\textrm{B}}^{\ast}$ being the effective Bohr radius and
$m_{\textrm{e}}$ being the bare electron mass, $\beta_{1}=2\pi\left[(\nu u\left(\nu\right))^{\prime\prime}-(\nu u\left(\nu\right))^{\prime}/2\nu\right]$,
$\beta_{2}=-2\pi\left[(\nu u\left(\nu\right))^{\prime\prime}\right]$.
At $\nu\approx1/2$, we have $\beta_{1}\approx2.3$ and $\beta_{2}\approx1.6$.

To observe the asymmetry inverse predicted in Eq.$\,$(\ref{eq:condition2}),
we require $\left|\alpha\right|\ll1$. It is not difficult to fulfill
the requirement in a GaAs-based 2DES, for which $a_{\textrm{c}}\approx62\,\textrm{nm}$.
For the experimental parameters of Ref.$\,$\citep{kamburov_what_2014},
i.e., $a=200\,\textrm{nm}$ and $B=14\,\textrm{T}$, the value of
$\alpha$ is $0.24$, fulfilling the requirement. In the strong field
limit $B\rightarrow\infty$, $l_{B}\rightarrow0$, we have $\alpha=a_{\textrm{c}}/a$.
Therefore, one can always fulfill the requirement by choosing a modulation
period $a$ much larger than $62\,\textrm{nm}$. We further note that
the modulation of the external magnetic field had already been achieved
for electrons by placing a ferromagnet or superconductor microstructure
on top of a 2DES$\,$\citep{ye_electrons_1995,izawa_magnetoresistance_1995,carmona_two_1995}.
We expect that similar techniques can be implemented for CF systems.

\section{Discussions and summary\label{sec:Discussions-and-summary}}

A natural question is what the Dirac CF theory would predict for the
asymmetry. Dirac CF theory proposed by Son also captures the effect
of the $\pi$-Berry phase, with a Berry curvature singularly distributed
in the center of the momentum space$\,$\citep{son_is_2015}. In Refs.$\,$\citep{cheung_weiss_2017,mitra_fluctuations_2019},
Cheung et al. conclude that for Dirac CFs, the difference between
the scalar potential modulation and the magnetic field modulation
is in the factor $\gamma$, i.e., $\gamma=1/4$ for the scalar potential
modulation and $\gamma=-1/4$ for the magnetic field modulation. It
would predict the interchange of the positions of the magnetoresistance
minimum and maximum. This is different from our prediction of the
inverse of the asymmetry. Our prediction is based on the dipole picture
of the ``subatomic'' structure of the CF, which is a result of the
microscopic Rezayi-Read wave function. However, the prediction for
Dirac CFs is based upon an effective field theory. Unfortunately,
for the Dirac CF theory, there is still no consensus on the microscopic
wave function and the ``subatomic'' structure.%

In summary, we theoretically study the manifestations of the uniform-Berry-curvature
picture in the geometrical resonance experiments for CFs. We show
that the modulation of an externally applied magnetic field will induce
an asymmetry opposite to that induced by a periodic scalar potential
modulation. This experiment can serve as a critical test to the uniform-Berry-curvature
picture. Since the effect originates from the dipole structure of
CFs, its successful observation will also provide an experimental
confirmation to the dipole picture of CFs initially proposed by Read.
\begin{acknowledgments}
This work is supported by National Basic Research Program of China
(973 Program) Grant No. 2015CB921101 and National Science Foundation
of China Grant No. 11325416.
\end{acknowledgments}

\bibliographystyle{apsrev4-1}
\bibliography{composite_fermion}

\begin{thebibliography}{39}%
\makeatletter
\providecommand \@ifxundefined [1]{%
 \@ifx{#1\undefined}
}%
\providecommand \@ifnum [1]{%
 \ifnum #1\expandafter \@firstoftwo
 \else \expandafter \@secondoftwo
 \fi
}%
\providecommand \@ifx [1]{%
 \ifx #1\expandafter \@firstoftwo
 \else \expandafter \@secondoftwo
 \fi
}%
\providecommand \natexlab [1]{#1}%
\providecommand \enquote  [1]{``#1''}%
\providecommand \bibnamefont  [1]{#1}%
\providecommand \bibfnamefont [1]{#1}%
\providecommand \citenamefont [1]{#1}%
\providecommand \href@noop [0]{\@secondoftwo}%
\providecommand \href [0]{\begingroup \@sanitize@url \@href}%
\providecommand \@href[1]{\@@startlink{#1}\@@href}%
\providecommand \@@href[1]{\endgroup#1\@@endlink}%
\providecommand \@sanitize@url [0]{\catcode `\\12\catcode `\$12\catcode
  `\&12\catcode `\#12\catcode `\^12\catcode `\_12\catcode `\%12\relax}%
\providecommand \@@startlink[1]{}%
\providecommand \@@endlink[0]{}%
\providecommand \url  [0]{\begingroup\@sanitize@url \@url }%
\providecommand \@url [1]{\endgroup\@href {#1}{\urlprefix }}%
\providecommand \urlprefix  [0]{URL }%
\providecommand \Eprint [0]{\href }%
\providecommand \doibase [0]{http://dx.doi.org/}%
\providecommand \selectlanguage [0]{\@gobble}%
\providecommand \bibinfo  [0]{\@secondoftwo}%
\providecommand \bibfield  [0]{\@secondoftwo}%
\providecommand \translation [1]{[#1]}%
\providecommand \BibitemOpen [0]{}%
\providecommand \bibitemStop [0]{}%
\providecommand \bibitemNoStop [0]{.\EOS\space}%
\providecommand \EOS [0]{\spacefactor3000\relax}%
\providecommand \BibitemShut  [1]{\csname bibitem#1\endcsname}%
\let\auto@bib@innerbib\@empty
\bibitem [{\citenamefont {Tsui}\ \emph {et~al.}(1982)\citenamefont {Tsui},
  \citenamefont {Stormer},\ and\ \citenamefont
  {Gossard}}]{tsui_two-dimensional_1982}%
  \BibitemOpen
  \bibfield  {author} {\bibinfo {author} {\bibfnamefont {D.~C.}\ \bibnamefont
  {Tsui}}, \bibinfo {author} {\bibfnamefont {H.~L.}\ \bibnamefont {Stormer}}, \
  and\ \bibinfo {author} {\bibfnamefont {A.~C.}\ \bibnamefont {Gossard}},\
  }\href {\doibase 10.1103/PhysRevLett.48.1559} {\bibfield  {journal} {\bibinfo
   {journal} {Phys. Rev. Lett.}\ }\textbf {\bibinfo {volume} {48}},\ \bibinfo
  {pages} {1559} (\bibinfo {year} {1982})}\BibitemShut {NoStop}%
\bibitem [{\citenamefont {Jain}(1989)}]{jain_composite-fermion_1989}%
  \BibitemOpen
  \bibfield  {author} {\bibinfo {author} {\bibfnamefont {J.~K.}\ \bibnamefont
  {Jain}},\ }\href {\doibase 10.1103/PhysRevLett.63.199} {\bibfield  {journal}
  {\bibinfo  {journal} {Phys. Rev. Lett.}\ }\textbf {\bibinfo {volume} {63}},\
  \bibinfo {pages} {199} (\bibinfo {year} {1989})}\BibitemShut {NoStop}%
\bibitem [{\citenamefont {Kalmeyer}\ and\ \citenamefont
  {Zhang}(1992)}]{kalmeyer_metallic_1992}%
  \BibitemOpen
  \bibfield  {author} {\bibinfo {author} {\bibfnamefont {V.}~\bibnamefont
  {Kalmeyer}}\ and\ \bibinfo {author} {\bibfnamefont {S.-C.}\ \bibnamefont
  {Zhang}},\ }\href {\doibase 10.1103/PhysRevB.46.9889} {\bibfield  {journal}
  {\bibinfo  {journal} {Phys. Rev. B}\ }\textbf {\bibinfo {volume} {46}},\
  \bibinfo {pages} {9889} (\bibinfo {year} {1992})}\BibitemShut {NoStop}%
\bibitem [{\citenamefont {Halperin}\ \emph {et~al.}(1993)\citenamefont
  {Halperin}, \citenamefont {Lee},\ and\ \citenamefont
  {Read}}]{halperin_theory_1993}%
  \BibitemOpen
  \bibfield  {author} {\bibinfo {author} {\bibfnamefont {B.~I.}\ \bibnamefont
  {Halperin}}, \bibinfo {author} {\bibfnamefont {P.~A.}\ \bibnamefont {Lee}}, \
  and\ \bibinfo {author} {\bibfnamefont {N.}~\bibnamefont {Read}},\ }\href
  {\doibase 10.1103/PhysRevB.47.7312} {\bibfield  {journal} {\bibinfo
  {journal} {Phys. Rev. B}\ }\textbf {\bibinfo {volume} {47}},\ \bibinfo
  {pages} {7312} (\bibinfo {year} {1993})}\BibitemShut {NoStop}%
\bibitem [{\citenamefont {Jain}(2007)}]{jain2007composite}%
  \BibitemOpen
  \bibfield  {author} {\bibinfo {author} {\bibfnamefont {J.}~\bibnamefont
  {Jain}},\ }\href {https://books.google.com.hk/books?id=0jv9UF6UL20C} {\emph
  {\bibinfo {title} {Composite Fermions}}}\ (\bibinfo  {publisher} {Cambridge
  University Press},\ \bibinfo {year} {2007})\BibitemShut {NoStop}%
\bibitem [{\citenamefont {Heinonen}(1998)}]{Heinonen}%
  \BibitemOpen
  \bibfield  {author} {\bibinfo {author} {\bibfnamefont {O.}~\bibnamefont
  {Heinonen}},\ }\href {\doibase 10.1142/3894} {\emph {\bibinfo {title}
  {Composite Fermions: A Unified View of the Quantum Hall Regime}}}\ (\bibinfo
  {publisher} {World Scientific},\ \bibinfo {year} {1998})\BibitemShut
  {NoStop}%
\bibitem [{\citenamefont {Kivelson}\ \emph {et~al.}(1997)\citenamefont
  {Kivelson}, \citenamefont {Lee}, \citenamefont {Krotov},\ and\ \citenamefont
  {Gan}}]{kivelson_composite-fermion_1997}%
  \BibitemOpen
  \bibfield  {author} {\bibinfo {author} {\bibfnamefont {S.~A.}\ \bibnamefont
  {Kivelson}}, \bibinfo {author} {\bibfnamefont {D.-H.}\ \bibnamefont {Lee}},
  \bibinfo {author} {\bibfnamefont {Y.}~\bibnamefont {Krotov}}, \ and\ \bibinfo
  {author} {\bibfnamefont {J.}~\bibnamefont {Gan}},\ }\href {\doibase
  10.1103/PhysRevB.55.15552} {\bibfield  {journal} {\bibinfo  {journal} {Phys.
  Rev. B}\ }\textbf {\bibinfo {volume} {55}},\ \bibinfo {pages} {15552}
  (\bibinfo {year} {1997})}\BibitemShut {NoStop}%
\bibitem [{\citenamefont {Son}(2015)}]{son_is_2015}%
  \BibitemOpen
  \bibfield  {author} {\bibinfo {author} {\bibfnamefont {D.~T.}\ \bibnamefont
  {Son}},\ }\href {\doibase 10.1103/PhysRevX.5.031027} {\bibfield  {journal}
  {\bibinfo  {journal} {Phys. Rev. X}\ }\textbf {\bibinfo {volume} {5}},\
  \bibinfo {pages} {031027} (\bibinfo {year} {2015})}\BibitemShut {NoStop}%
\bibitem [{\citenamefont {Balram}\ and\ \citenamefont
  {Jain}(2016)}]{balram_nature_2016}%
  \BibitemOpen
  \bibfield  {author} {\bibinfo {author} {\bibfnamefont {A.~C.}\ \bibnamefont
  {Balram}}\ and\ \bibinfo {author} {\bibfnamefont {J.~K.}\ \bibnamefont
  {Jain}},\ }\href {\doibase 10.1103/PhysRevB.93.235152} {\bibfield  {journal}
  {\bibinfo  {journal} {Phys. Rev. B}\ }\textbf {\bibinfo {volume} {93}},\
  \bibinfo {pages} {235152} (\bibinfo {year} {2016})}\BibitemShut {NoStop}%
\bibitem [{\citenamefont {Shi}\ and\ \citenamefont
  {Ji}(2018)}]{shi_dynamics_2018}%
  \BibitemOpen
  \bibfield  {author} {\bibinfo {author} {\bibfnamefont {J.}~\bibnamefont
  {Shi}}\ and\ \bibinfo {author} {\bibfnamefont {W.}~\bibnamefont {Ji}},\
  }\href {\doibase 10.1103/PhysRevB.97.125133} {\bibfield  {journal} {\bibinfo
  {journal} {Phys. Rev. B}\ }\textbf {\bibinfo {volume} {97}},\ \bibinfo
  {pages} {125133} (\bibinfo {year} {2018})}\BibitemShut {NoStop}%
\bibitem [{\citenamefont {Shi}()}]{shi2017chernsimons}%
  \BibitemOpen
  \bibfield  {author} {\bibinfo {author} {\bibfnamefont {J.}~\bibnamefont
  {Shi}},\ }\href@noop {} {}\Eprint {http://arxiv.org/abs/1704.07712}
  {arXiv:1704.07712} \BibitemShut {NoStop}%
\bibitem [{\citenamefont {Geraedts}\ \emph {et~al.}(2018)\citenamefont
  {Geraedts}, \citenamefont {Wang}, \citenamefont {Rezayi},\ and\ \citenamefont
  {Haldane}}]{geraedts_berry_2018}%
  \BibitemOpen
  \bibfield  {author} {\bibinfo {author} {\bibfnamefont {S.~D.}\ \bibnamefont
  {Geraedts}}, \bibinfo {author} {\bibfnamefont {J.}~\bibnamefont {Wang}},
  \bibinfo {author} {\bibfnamefont {E.}~\bibnamefont {Rezayi}}, \ and\ \bibinfo
  {author} {\bibfnamefont {F.}~\bibnamefont {Haldane}},\ }\href {\doibase
  10.1103/PhysRevLett.121.147202} {\bibfield  {journal} {\bibinfo  {journal}
  {Phys. Rev. Lett.}\ }\textbf {\bibinfo {volume} {121}},\ \bibinfo {pages}
  {147202} (\bibinfo {year} {2018})}\BibitemShut {NoStop}%
\bibitem [{\citenamefont {Ji}\ and\ \citenamefont {Shi}()}]{ji2019berry}%
  \BibitemOpen
  \bibfield  {author} {\bibinfo {author} {\bibfnamefont {G.}~\bibnamefont
  {Ji}}\ and\ \bibinfo {author} {\bibfnamefont {J.}~\bibnamefont {Shi}},\
  }\href@noop {} {}\Eprint {http://arxiv.org/abs/1901.00321} {arXiv:1901.00321}
  \BibitemShut {NoStop}%
\bibitem [{\citenamefont {Rezayi}\ and\ \citenamefont
  {Read}(1994)}]{rezayi_fermi-liquid-like_1994}%
  \BibitemOpen
  \bibfield  {author} {\bibinfo {author} {\bibfnamefont {E.}~\bibnamefont
  {Rezayi}}\ and\ \bibinfo {author} {\bibfnamefont {N.}~\bibnamefont {Read}},\
  }\href {\doibase 10.1103/PhysRevLett.72.900} {\bibfield  {journal} {\bibinfo
  {journal} {Phys. Rev. Lett.}\ }\textbf {\bibinfo {volume} {72}},\ \bibinfo
  {pages} {900} (\bibinfo {year} {1994})}\BibitemShut {NoStop}%
\bibitem [{\citenamefont {Geraedts}\ \emph {et~al.}(2016)\citenamefont
  {Geraedts}, \citenamefont {Zaletel}, \citenamefont {Mong}, \citenamefont
  {Metlitski}, \citenamefont {Vishwanath},\ and\ \citenamefont
  {Motrunich}}]{geraedts_half-filled_2016}%
  \BibitemOpen
  \bibfield  {author} {\bibinfo {author} {\bibfnamefont {S.~D.}\ \bibnamefont
  {Geraedts}}, \bibinfo {author} {\bibfnamefont {M.~P.}\ \bibnamefont
  {Zaletel}}, \bibinfo {author} {\bibfnamefont {R.~S.~K.}\ \bibnamefont
  {Mong}}, \bibinfo {author} {\bibfnamefont {M.~A.}\ \bibnamefont {Metlitski}},
  \bibinfo {author} {\bibfnamefont {A.}~\bibnamefont {Vishwanath}}, \ and\
  \bibinfo {author} {\bibfnamefont {O.~I.}\ \bibnamefont {Motrunich}},\ }\href
  {\doibase 10.1126/science.aad4302} {\bibfield  {journal} {\bibinfo  {journal}
  {Science}\ }\textbf {\bibinfo {volume} {352}},\ \bibinfo {pages} {197}
  (\bibinfo {year} {2016})}\BibitemShut {NoStop}%
\bibitem [{\citenamefont {Pan}\ \emph {et~al.}(2017)\citenamefont {Pan},
  \citenamefont {Kang}, \citenamefont {Baldwin}, \citenamefont {West},
  \citenamefont {Pfeiffer},\ and\ \citenamefont {Tsui}}]{pan_berry_2017}%
  \BibitemOpen
  \bibfield  {author} {\bibinfo {author} {\bibfnamefont {W.}~\bibnamefont
  {Pan}}, \bibinfo {author} {\bibfnamefont {W.}~\bibnamefont {Kang}}, \bibinfo
  {author} {\bibfnamefont {K.~W.}\ \bibnamefont {Baldwin}}, \bibinfo {author}
  {\bibfnamefont {K.~W.}\ \bibnamefont {West}}, \bibinfo {author}
  {\bibfnamefont {L.~N.}\ \bibnamefont {Pfeiffer}}, \ and\ \bibinfo {author}
  {\bibfnamefont {D.~C.}\ \bibnamefont {Tsui}},\ }\href {\doibase
  10.1038/nphys4231} {\bibfield  {journal} {\bibinfo  {journal} {Nature
  Physics}\ }\textbf {\bibinfo {volume} {13}},\ \bibinfo {pages} {1168}
  (\bibinfo {year} {2017})}\BibitemShut {NoStop}%
\bibitem [{\citenamefont {Kamburov}\ \emph {et~al.}(2014)\citenamefont
  {Kamburov}, \citenamefont {Liu}, \citenamefont {Mueed}, \citenamefont
  {Shayegan}, \citenamefont {Pfeiffer}, \citenamefont {West},\ and\
  \citenamefont {Baldwin}}]{kamburov_what_2014}%
  \BibitemOpen
  \bibfield  {author} {\bibinfo {author} {\bibfnamefont {D.}~\bibnamefont
  {Kamburov}}, \bibinfo {author} {\bibfnamefont {Y.}~\bibnamefont {Liu}},
  \bibinfo {author} {\bibfnamefont {M.}~\bibnamefont {Mueed}}, \bibinfo
  {author} {\bibfnamefont {M.}~\bibnamefont {Shayegan}}, \bibinfo {author}
  {\bibfnamefont {L.}~\bibnamefont {Pfeiffer}}, \bibinfo {author}
  {\bibfnamefont {K.}~\bibnamefont {West}}, \ and\ \bibinfo {author}
  {\bibfnamefont {K.}~\bibnamefont {Baldwin}},\ }\href {\doibase
  10.1103/PhysRevLett.113.196801} {\bibfield  {journal} {\bibinfo  {journal}
  {Phys. Rev. Lett.}\ }\textbf {\bibinfo {volume} {113}},\ \bibinfo {pages}
  {196801} (\bibinfo {year} {2014})}\BibitemShut {NoStop}%
\bibitem [{\citenamefont {Read}(1994)}]{read_theory_1994}%
  \BibitemOpen
  \bibfield  {author} {\bibinfo {author} {\bibfnamefont {N.}~\bibnamefont
  {Read}},\ }\href {\doibase 10.1088/0268-1242/9/11S/002} {\bibfield  {journal}
  {\bibinfo  {journal} {Semicond. Sci. Technol.}\ }\textbf {\bibinfo {volume}
  {9}},\ \bibinfo {pages} {1859} (\bibinfo {year} {1994})}\BibitemShut
  {NoStop}%
\bibitem [{\citenamefont {Read}(1996)}]{read_recent_1996}%
  \BibitemOpen
  \bibfield  {author} {\bibinfo {author} {\bibfnamefont {N.}~\bibnamefont
  {Read}},\ }\href {\doibase 10.1016/0039-6028(96)00318-4} {\bibfield
  {journal} {\bibinfo  {journal} {Surface Science}\ }\textbf {\bibinfo {volume}
  {361-362}},\ \bibinfo {pages} {7} (\bibinfo {year} {1996})}\BibitemShut
  {NoStop}%
\bibitem [{\citenamefont {Xiao}\ \emph {et~al.}(2005)\citenamefont {Xiao},
  \citenamefont {Shi},\ and\ \citenamefont {Niu}}]{xiao_berry_2005}%
  \BibitemOpen
  \bibfield  {author} {\bibinfo {author} {\bibfnamefont {D.}~\bibnamefont
  {Xiao}}, \bibinfo {author} {\bibfnamefont {J.}~\bibnamefont {Shi}}, \ and\
  \bibinfo {author} {\bibfnamefont {Q.}~\bibnamefont {Niu}},\ }\href {\doibase
  10.1103/PhysRevLett.95.137204} {\bibfield  {journal} {\bibinfo  {journal}
  {Phys. Rev. Lett.}\ }\textbf {\bibinfo {volume} {95}},\ \bibinfo {pages}
  {137204} (\bibinfo {year} {2005})}\BibitemShut {NoStop}%
\bibitem [{\citenamefont {Weiss}\ \emph {et~al.}(1989)\citenamefont {Weiss},
  \citenamefont {Klitzing}, \citenamefont {Ploog},\ and\ \citenamefont
  {Weimann}}]{weiss_magnetoresistance_1989}%
  \BibitemOpen
  \bibfield  {author} {\bibinfo {author} {\bibfnamefont {D.}~\bibnamefont
  {Weiss}}, \bibinfo {author} {\bibfnamefont {K.~V.}\ \bibnamefont {Klitzing}},
  \bibinfo {author} {\bibfnamefont {K.}~\bibnamefont {Ploog}}, \ and\ \bibinfo
  {author} {\bibfnamefont {G.}~\bibnamefont {Weimann}},\ }\href {\doibase
  10.1209/0295-5075/8/2/012} {\bibfield  {journal} {\bibinfo  {journal} {EPL}\
  }\textbf {\bibinfo {volume} {8}},\ \bibinfo {pages} {179} (\bibinfo {year}
  {1989})}\BibitemShut {NoStop}%
\bibitem [{\citenamefont {Willett}\ \emph {et~al.}(1993)\citenamefont
  {Willett}, \citenamefont {Ruel}, \citenamefont {West},\ and\ \citenamefont
  {Pfeiffer}}]{willett_experimental_1993}%
  \BibitemOpen
  \bibfield  {author} {\bibinfo {author} {\bibfnamefont {R.~L.}\ \bibnamefont
  {Willett}}, \bibinfo {author} {\bibfnamefont {R.~R.}\ \bibnamefont {Ruel}},
  \bibinfo {author} {\bibfnamefont {K.~W.}\ \bibnamefont {West}}, \ and\
  \bibinfo {author} {\bibfnamefont {L.~N.}\ \bibnamefont {Pfeiffer}},\ }\href
  {\doibase 10.1103/PhysRevLett.71.3846} {\bibfield  {journal} {\bibinfo
  {journal} {Phys. Rev. Lett.}\ }\textbf {\bibinfo {volume} {71}},\ \bibinfo
  {pages} {3846} (\bibinfo {year} {1993})}\BibitemShut {NoStop}%
\bibitem [{\citenamefont {Kang}\ \emph {et~al.}(1993)\citenamefont {Kang},
  \citenamefont {Stormer}, \citenamefont {Pfeiffer}, \citenamefont {Baldwin},\
  and\ \citenamefont {West}}]{kang_how_1993}%
  \BibitemOpen
  \bibfield  {author} {\bibinfo {author} {\bibfnamefont {W.}~\bibnamefont
  {Kang}}, \bibinfo {author} {\bibfnamefont {H.~L.}\ \bibnamefont {Stormer}},
  \bibinfo {author} {\bibfnamefont {L.~N.}\ \bibnamefont {Pfeiffer}}, \bibinfo
  {author} {\bibfnamefont {K.~W.}\ \bibnamefont {Baldwin}}, \ and\ \bibinfo
  {author} {\bibfnamefont {K.~W.}\ \bibnamefont {West}},\ }\href {\doibase
  10.1103/PhysRevLett.71.3850} {\bibfield  {journal} {\bibinfo  {journal}
  {Phys. Rev. Lett.}\ }\textbf {\bibinfo {volume} {71}},\ \bibinfo {pages}
  {3850} (\bibinfo {year} {1993})}\BibitemShut {NoStop}%
\bibitem [{\citenamefont {Goldman}\ \emph {et~al.}(1994)\citenamefont
  {Goldman}, \citenamefont {Su},\ and\ \citenamefont
  {Jain}}]{goldman_detection_1994}%
  \BibitemOpen
  \bibfield  {author} {\bibinfo {author} {\bibfnamefont {V.~J.}\ \bibnamefont
  {Goldman}}, \bibinfo {author} {\bibfnamefont {B.}~\bibnamefont {Su}}, \ and\
  \bibinfo {author} {\bibfnamefont {J.~K.}\ \bibnamefont {Jain}},\ }\href
  {\doibase 10.1103/PhysRevLett.72.2065} {\bibfield  {journal} {\bibinfo
  {journal} {Phys. Rev. Lett.}\ }\textbf {\bibinfo {volume} {72}},\ \bibinfo
  {pages} {2065} (\bibinfo {year} {1994})}\BibitemShut {NoStop}%
\bibitem [{\citenamefont {Smet}\ \emph {et~al.}(1996)\citenamefont {Smet},
  \citenamefont {Weiss}, \citenamefont {Blick}, \citenamefont {L\"{u}tjering},
  \citenamefont {von Klitzing}, \citenamefont {Fleischmann}, \citenamefont
  {Ketzmerick}, \citenamefont {Geisel},\ and\ \citenamefont
  {Weimann}}]{smet_magnetic_1996}%
  \BibitemOpen
  \bibfield  {author} {\bibinfo {author} {\bibfnamefont {J.~H.}\ \bibnamefont
  {Smet}}, \bibinfo {author} {\bibfnamefont {D.}~\bibnamefont {Weiss}},
  \bibinfo {author} {\bibfnamefont {R.~H.}\ \bibnamefont {Blick}}, \bibinfo
  {author} {\bibfnamefont {G.}~\bibnamefont {L\"{u}tjering}}, \bibinfo {author}
  {\bibfnamefont {K.}~\bibnamefont {von Klitzing}}, \bibinfo {author}
  {\bibfnamefont {R.}~\bibnamefont {Fleischmann}}, \bibinfo {author}
  {\bibfnamefont {R.}~\bibnamefont {Ketzmerick}}, \bibinfo {author}
  {\bibfnamefont {T.}~\bibnamefont {Geisel}}, \ and\ \bibinfo {author}
  {\bibfnamefont {G.}~\bibnamefont {Weimann}},\ }\href {\doibase
  10.1103/PhysRevLett.77.2272} {\bibfield  {journal} {\bibinfo  {journal}
  {Phys. Rev. Lett.}\ }\textbf {\bibinfo {volume} {77}},\ \bibinfo {pages}
  {2272} (\bibinfo {year} {1996})}\BibitemShut {NoStop}%
\bibitem [{\citenamefont {Skuras}\ \emph {et~al.}(1997)\citenamefont {Skuras},
  \citenamefont {Long}, \citenamefont {Larkin}, \citenamefont {Davies},\ and\
  \citenamefont {Holland}}]{skuras_anisotropic_1997}%
  \BibitemOpen
  \bibfield  {author} {\bibinfo {author} {\bibfnamefont {E.}~\bibnamefont
  {Skuras}}, \bibinfo {author} {\bibfnamefont {A.~R.}\ \bibnamefont {Long}},
  \bibinfo {author} {\bibfnamefont {I.~A.}\ \bibnamefont {Larkin}}, \bibinfo
  {author} {\bibfnamefont {J.~H.}\ \bibnamefont {Davies}}, \ and\ \bibinfo
  {author} {\bibfnamefont {M.~C.}\ \bibnamefont {Holland}},\ }\href {\doibase
  10.1063/1.118301} {\bibfield  {journal} {\bibinfo  {journal} {Appl. Phys.
  Lett.}\ }\textbf {\bibinfo {volume} {70}},\ \bibinfo {pages} {871} (\bibinfo
  {year} {1997})}\BibitemShut {NoStop}%
\bibitem [{\citenamefont {Peeters}\ and\ \citenamefont
  {Vasilopoulos}(1993)}]{peeters_quantum_1993}%
  \BibitemOpen
  \bibfield  {author} {\bibinfo {author} {\bibfnamefont {F.~M.}\ \bibnamefont
  {Peeters}}\ and\ \bibinfo {author} {\bibfnamefont {P.}~\bibnamefont
  {Vasilopoulos}},\ }\href {\doibase 10.1103/PhysRevB.47.1466} {\bibfield
  {journal} {\bibinfo  {journal} {Phys. Rev. B}\ }\textbf {\bibinfo {volume}
  {47}},\ \bibinfo {pages} {1466} (\bibinfo {year} {1993})}\BibitemShut
  {NoStop}%
\bibitem [{\citenamefont
  {Beenakker}(1989)}]{beenakker_guiding-center-drift_1989}%
  \BibitemOpen
  \bibfield  {author} {\bibinfo {author} {\bibfnamefont {C.~W.~J.}\
  \bibnamefont {Beenakker}},\ }\href {\doibase 10.1103/PhysRevLett.62.2020}
  {\bibfield  {journal} {\bibinfo  {journal} {Phys. Rev. Lett.}\ }\textbf
  {\bibinfo {volume} {62}},\ \bibinfo {pages} {2020} (\bibinfo {year}
  {1989})}\BibitemShut {NoStop}%
\bibitem [{\citenamefont {Peeters}\ and\ \citenamefont
  {Vasilopoulos}(1992)}]{peeters_electrical_1992}%
  \BibitemOpen
  \bibfield  {author} {\bibinfo {author} {\bibfnamefont {F.~M.}\ \bibnamefont
  {Peeters}}\ and\ \bibinfo {author} {\bibfnamefont {P.}~\bibnamefont
  {Vasilopoulos}},\ }\href {\doibase 10.1103/PhysRevB.46.4667} {\bibfield
  {journal} {\bibinfo  {journal} {Phys. Rev. B}\ }\textbf {\bibinfo {volume}
  {46}},\ \bibinfo {pages} {4667} (\bibinfo {year} {1992})}\BibitemShut
  {NoStop}%
\bibitem [{\citenamefont {Simon}\ \emph {et~al.}(1996)\citenamefont {Simon},
  \citenamefont {Stern},\ and\ \citenamefont
  {Halperin}}]{simon_composite_1996}%
  \BibitemOpen
  \bibfield  {author} {\bibinfo {author} {\bibfnamefont {S.~H.}\ \bibnamefont
  {Simon}}, \bibinfo {author} {\bibfnamefont {A.}~\bibnamefont {Stern}}, \ and\
  \bibinfo {author} {\bibfnamefont {B.~I.}\ \bibnamefont {Halperin}},\ }\href
  {\doibase 10.1103/PhysRevB.54.R11114} {\bibfield  {journal} {\bibinfo
  {journal} {Phys. Rev. B}\ }\textbf {\bibinfo {volume} {54}},\ \bibinfo
  {pages} {R11114} (\bibinfo {year} {1996})}\BibitemShut {NoStop}%
\bibitem [{\citenamefont {Ferconi}\ \emph {et~al.}(1995)\citenamefont
  {Ferconi}, \citenamefont {Geller},\ and\ \citenamefont
  {Vignale}}]{ferconi_edge_1995}%
  \BibitemOpen
  \bibfield  {author} {\bibinfo {author} {\bibfnamefont {M.}~\bibnamefont
  {Ferconi}}, \bibinfo {author} {\bibfnamefont {M.~R.}\ \bibnamefont {Geller}},
  \ and\ \bibinfo {author} {\bibfnamefont {G.}~\bibnamefont {Vignale}},\ }\href
  {\doibase 10.1103/PhysRevB.52.16357} {\bibfield  {journal} {\bibinfo
  {journal} {Phys. Rev. B}\ }\textbf {\bibinfo {volume} {52}},\ \bibinfo
  {pages} {16357} (\bibinfo {year} {1995})}\BibitemShut {NoStop}%
\bibitem [{\citenamefont {Heinonen}\ \emph {et~al.}(1995)\citenamefont
  {Heinonen}, \citenamefont {Lubin},\ and\ \citenamefont
  {Johnson}}]{heinonen_ensemble_1995}%
  \BibitemOpen
  \bibfield  {author} {\bibinfo {author} {\bibfnamefont {O.}~\bibnamefont
  {Heinonen}}, \bibinfo {author} {\bibfnamefont {M.~I.}\ \bibnamefont {Lubin}},
  \ and\ \bibinfo {author} {\bibfnamefont {M.~D.}\ \bibnamefont {Johnson}},\
  }\href {\doibase 10.1103/PhysRevLett.75.4110} {\bibfield  {journal} {\bibinfo
   {journal} {Phys. Rev. Lett.}\ }\textbf {\bibinfo {volume} {75}},\ \bibinfo
  {pages} {4110} (\bibinfo {year} {1995})}\BibitemShut {NoStop}%
\bibitem [{\citenamefont {Zhao}\ \emph {et~al.}(2017)\citenamefont {Zhao},
  \citenamefont {Thakurathi}, \citenamefont {Jain}, \citenamefont {Sen},\ and\
  \citenamefont {Jain}}]{zhao_density-functional_2017}%
  \BibitemOpen
  \bibfield  {author} {\bibinfo {author} {\bibfnamefont {J.}~\bibnamefont
  {Zhao}}, \bibinfo {author} {\bibfnamefont {M.}~\bibnamefont {Thakurathi}},
  \bibinfo {author} {\bibfnamefont {M.}~\bibnamefont {Jain}}, \bibinfo {author}
  {\bibfnamefont {D.}~\bibnamefont {Sen}}, \ and\ \bibinfo {author}
  {\bibfnamefont {J.}~\bibnamefont {Jain}},\ }\href {\doibase
  10.1103/PhysRevLett.118.196802} {\bibfield  {journal} {\bibinfo  {journal}
  {Phys. Rev. Lett.}\ }\textbf {\bibinfo {volume} {118}},\ \bibinfo {pages}
  {196802} (\bibinfo {year} {2017})}\BibitemShut {NoStop}%
\bibitem [{\citenamefont {Fano}\ and\ \citenamefont
  {Ortolani}(1988)}]{fano_interpolation_1988}%
  \BibitemOpen
  \bibfield  {author} {\bibinfo {author} {\bibfnamefont {G.}~\bibnamefont
  {Fano}}\ and\ \bibinfo {author} {\bibfnamefont {F.}~\bibnamefont
  {Ortolani}},\ }\href {\doibase 10.1103/PhysRevB.37.8179} {\bibfield
  {journal} {\bibinfo  {journal} {Phys. Rev. B}\ }\textbf {\bibinfo {volume}
  {37}},\ \bibinfo {pages} {8179} (\bibinfo {year} {1988})}\BibitemShut
  {NoStop}%
\bibitem [{\citenamefont {Ye}\ \emph {et~al.}(1995)\citenamefont {Ye},
  \citenamefont {Weiss}, \citenamefont {Gerhardts}, \citenamefont {Seeger},
  \citenamefont {von Klitzing}, \citenamefont {Eberl},\ and\ \citenamefont
  {Nickel}}]{ye_electrons_1995}%
  \BibitemOpen
  \bibfield  {author} {\bibinfo {author} {\bibfnamefont {P.~D.}\ \bibnamefont
  {Ye}}, \bibinfo {author} {\bibfnamefont {D.}~\bibnamefont {Weiss}}, \bibinfo
  {author} {\bibfnamefont {R.~R.}\ \bibnamefont {Gerhardts}}, \bibinfo {author}
  {\bibfnamefont {M.}~\bibnamefont {Seeger}}, \bibinfo {author} {\bibfnamefont
  {K.}~\bibnamefont {von Klitzing}}, \bibinfo {author} {\bibfnamefont
  {K.}~\bibnamefont {Eberl}}, \ and\ \bibinfo {author} {\bibfnamefont
  {H.}~\bibnamefont {Nickel}},\ }\href {\doibase 10.1103/PhysRevLett.74.3013}
  {\bibfield  {journal} {\bibinfo  {journal} {Phys. Rev. Lett.}\ }\textbf
  {\bibinfo {volume} {74}},\ \bibinfo {pages} {3013} (\bibinfo {year}
  {1995})}\BibitemShut {NoStop}%
\bibitem [{\citenamefont {Izawa}\ \emph {et~al.}(1995)\citenamefont {Izawa},
  \citenamefont {Katsumoto}, \citenamefont {Endo},\ and\ \citenamefont
  {Iye}}]{izawa_magnetoresistance_1995}%
  \BibitemOpen
  \bibfield  {author} {\bibinfo {author} {\bibfnamefont {S.-i.}\ \bibnamefont
  {Izawa}}, \bibinfo {author} {\bibfnamefont {S.}~\bibnamefont {Katsumoto}},
  \bibinfo {author} {\bibfnamefont {A.}~\bibnamefont {Endo}}, \ and\ \bibinfo
  {author} {\bibfnamefont {Y.}~\bibnamefont {Iye}},\ }\href {\doibase
  10.1143/jpsj.64.706} {\bibfield  {journal} {\bibinfo  {journal} {Journal of
  the Physical Society of Japan}\ }\textbf {\bibinfo {volume} {64}},\ \bibinfo
  {pages} {706} (\bibinfo {year} {1995})}\BibitemShut {NoStop}%
\bibitem [{\citenamefont {Carmona}\ \emph {et~al.}(1995)\citenamefont
  {Carmona}, \citenamefont {Geim}, \citenamefont {Nogaret}, \citenamefont
  {Main}, \citenamefont {Foster}, \citenamefont {Henini}, \citenamefont
  {Beaumont},\ and\ \citenamefont {Blamire}}]{carmona_two_1995}%
  \BibitemOpen
  \bibfield  {author} {\bibinfo {author} {\bibfnamefont {H.~A.}\ \bibnamefont
  {Carmona}}, \bibinfo {author} {\bibfnamefont {A.~K.}\ \bibnamefont {Geim}},
  \bibinfo {author} {\bibfnamefont {A.}~\bibnamefont {Nogaret}}, \bibinfo
  {author} {\bibfnamefont {P.~C.}\ \bibnamefont {Main}}, \bibinfo {author}
  {\bibfnamefont {T.~J.}\ \bibnamefont {Foster}}, \bibinfo {author}
  {\bibfnamefont {M.}~\bibnamefont {Henini}}, \bibinfo {author} {\bibfnamefont
  {S.~P.}\ \bibnamefont {Beaumont}}, \ and\ \bibinfo {author} {\bibfnamefont
  {M.~G.}\ \bibnamefont {Blamire}},\ }\href {\doibase
  10.1103/PhysRevLett.74.3009} {\bibfield  {journal} {\bibinfo  {journal}
  {Phys. Rev. Lett.}\ }\textbf {\bibinfo {volume} {74}},\ \bibinfo {pages}
  {3009} (\bibinfo {year} {1995})}\BibitemShut {NoStop}%
\bibitem [{\citenamefont {Cheung}\ \emph {et~al.}(2017)\citenamefont {Cheung},
  \citenamefont {Raghu},\ and\ \citenamefont {Mulligan}}]{cheung_weiss_2017}%
  \BibitemOpen
  \bibfield  {author} {\bibinfo {author} {\bibfnamefont {A.~K.~C.}\
  \bibnamefont {Cheung}}, \bibinfo {author} {\bibfnamefont {S.}~\bibnamefont
  {Raghu}}, \ and\ \bibinfo {author} {\bibfnamefont {M.}~\bibnamefont
  {Mulligan}},\ }\href {\doibase 10.1103/PhysRevB.95.235424} {\bibfield
  {journal} {\bibinfo  {journal} {Phys. Rev. B}\ }\textbf {\bibinfo {volume}
  {95}},\ \bibinfo {pages} {235424} (\bibinfo {year} {2017})}\BibitemShut
  {NoStop}%
\bibitem [{\citenamefont {Mitra}\ and\ \citenamefont
  {Mulligan}(2019)}]{mitra_fluctuations_2019}%
  \BibitemOpen
  \bibfield  {author} {\bibinfo {author} {\bibfnamefont {A.}~\bibnamefont
  {Mitra}}\ and\ \bibinfo {author} {\bibfnamefont {M.}~\bibnamefont
  {Mulligan}},\ }\href {\doibase 10.1103/PhysRevB.100.165122} {\bibfield
  {journal} {\bibinfo  {journal} {Phys. Rev. B}\ }\textbf {\bibinfo {volume}
  {100}},\ \bibinfo {pages} {165122} (\bibinfo {year} {2019})}\BibitemShut
  {NoStop}%
\end{thebibliography}%

\end{document}